\documentclass[ajl]{emulateapj}

\usepackage{graphicx,epsfig,natbib,color}
\usepackage{lscape}
\usepackage{graphicx}
\usepackage{epsfig}
\usepackage{natbib}
\usepackage[colorlinks, citecolor=blue]{hyperref}
\usepackage{bm}

\slugcomment{}

\begin{document}
\title{A Two-ribbon White-light Flare Associated with a Failed Solar Eruption Observed by \textit{ONSET}, \textit{SDO}, and \textit{IRIS}}

\author{X. Cheng$^{1,2,3}$, Q. Hao$^{1,3}$, M. D. Ding$^{1,3}$, K. Liu$^{4}$, P. F. Chen$^{1,3}$, C. Fang$^{1,3}$, \& Y. D. Liu$^{2}$}

\affil{$^1$School of Astronomy and Space Science, Nanjing University, Nanjing 210093, China}\email{xincheng@nju.edu.cn}
\affil{$^2$State Key Laboratory of Space Weather, Chinese Academy of Sciences, Beijing 100190, China}
\affil{$^3$Key Laboratory for Modern Astronomy and Astrophysics (Nanjing University), Ministry of Education, Nanjing 210093, China}
\affil{$^4$Key Laboratory of Geospace Environment, University of Science and Technology of China, Hefei 230026, China}

\begin{abstract}
Two-ribbon brightenings are one of the most remarkable characteristics of an eruptive solar flare and are often used for predicting the occurrence of coronal mass ejections (CMEs). Nevertheless, it was called in question recently whether all two-ribbon flares are eruptive. In this paper, we investigate a two ribbon-like white-light (WL) flare that is associated with a failed magnetic flux rope (MFR) eruption on 2015 January 13, which has no accompanying CME in the WL coronagraph. Observations by \textit{Optical and Near-infrared Solar Eruption Tracer} and \textit{Solar Dynamics Observatory} reveal that, with the increase of the flare emission and the acceleration of the unsuccessfully erupting MFR, two isolated kernels appear at the WL 3600 {\AA} passband and quickly develop into two elongated ribbon-like structures. The evolution of the WL continuum enhancement is completely coincident in time with the variation of \textit{Fermi} hard X-ray 26--50 keV flux. Increase of continuum emission is also clearly visible at the whole FUV and NUV passbands observed by \textit{Interface Region Imaging Spectrograph}. Moreover, in one WL kernel, the \ion{Si}{4}, \ion{C}{2}, and \ion{Mg}{2} h/k lines display significant enhancement and non-thermal broadening. However, their Doppler velocity pattern is location-dependent. At the strongly bright pixels, these lines exhibit a blueshift; while at moderately bright ones, the lines are generally redshifted. These results show that the failed MFR eruption is also able to produce a two-ribbon flare and high-energy electrons that heat the lower atmosphere, causing the enhancement of the WL and FUV/NUV continuum  emissions and chromospheric evaporation.
\end{abstract}


\section{Introduction}
Solar flares are one of the most energetic phenomena that take place in the solar atmosphere. The induced emission displays a sudden and rapid increase over the whole electromagnetic spectrum from decameter radio waves to $\gamma$ rays at 100 MeV \citep{benz08}. In the standard flare model \citep{sturrock66,hirayama74,kopp76}, magnetic reconnection is believed to be a fundamental energy release mechanism and is capable of efficiently converting magnetic free energy stored in the corona into thermal energy, kinematic energy, and particles acceleration \citep{priest02}. 

Magnetic flux rope (MFR), a set of magnetic field lines wrapping around its central axis, is another important structural component in reconnection-involved eruption scenario \citep{cheng11_fluxrope}. This kind of helically coherent magnetic structure may exist prior to the eruption as seen in the high-temperature passbands of the Atmospheric Imaging Assembly \citep[AIA;][]{lemen12} on board \textit{Solar Dynamics Observatory} \citep[\textit{SDO};][]{zhang12,cheng13_driver,patsourakos13,lileping13,liting13,chenbin14,yang14,song15_firsttaste}. The rapid eruption of the MFR drives the formation of a coronal mass ejection (CME), facilitating and in turn being promoted by the fast flare reconnection \citep{cheng13_driver,cheng13_double}. The newly reconnected field lines initially manifest as a cusp-shaped structure and subsequently shrink into flare loops. During the flare process, electrons are also accelerated and then quickly stream down along the flare loops, heating the chromosphere and producing enhanced flare emission at the footpoints that form two elongated ribbons if the reconnection develops above a sequence of flare arcades \citep{forbes95}. 

\begin{table*}
\caption{Capabilities of instruments}\vspace{0.0\textwidth}
\label{tb1}
\resizebox{\textwidth}{!}{
%\begin{tabular}{lp{10cm}}\tableline \tableline
\begin{tabular}{ccccc}
\hline\hline
Telescope  & Spatial resolution & Cadence & Wavelength coverage & Spectral resolution  \\
                     & [\arcsec] & [s] & [{\AA}] & [km s${^{-1}}$]  \\
\hline                    
\textit{ONSET} &1.0 &20&10830, 6563, 3600, and 4250&- \\
\textit{SDO}-AIA &1.2&12&131, 94, 335, 193, 211, 171, 304, 1600, 1700, 4500&- \\
\textit{SDO}-HMI &1.0 &45&6173&- \\
\textit{IRIS} &0.4&10&1332-1358, 1389-1407, and 2783-2835&1 \\
\tableline
\vspace{0.01\textwidth}
\end{tabular}}
\end{table*}

If flares are strong enough, continuum emissions at the visible wavelengths may present an enhancement. This kind of flares is also called white-light flares \citep[WLFs;][]{svestka66,neidig93,fang95}. \citet{machado1986} classified all WLFs into two different types (I and II). For the type I WLFs, the WL emission coincides temporally with the peak of the hard X-ray (HXR) emission, and the Balmer lines are very broad \citep[e.g.,][]{fang95}; while for the type II WLFs, the situation is different \citep{ding99aa,ding99apj}. Recent studies further reveal that the WL brightening in the type I WLFs often appears as discrete kernels and displays the characteristics: (1) the locations of the WL kernels are cospatial with those of the HXR sources \citep{hudson92,metcalf03,chenqr05,chenqr06,jess08,krucker11,kerr14}, and (2) the source size of the WL kernels gradually decreases with the increase of the atmospheric depth \citep{xu12}. These results suggest that the WL emission enhancement of type I WLFs is associated with energetic electrons that heat the chromosphere directly and the photosphere indirectly through radiative backwarming \citep{ding03}.

\begin{figure}
\center {\includegraphics[width=8cm]{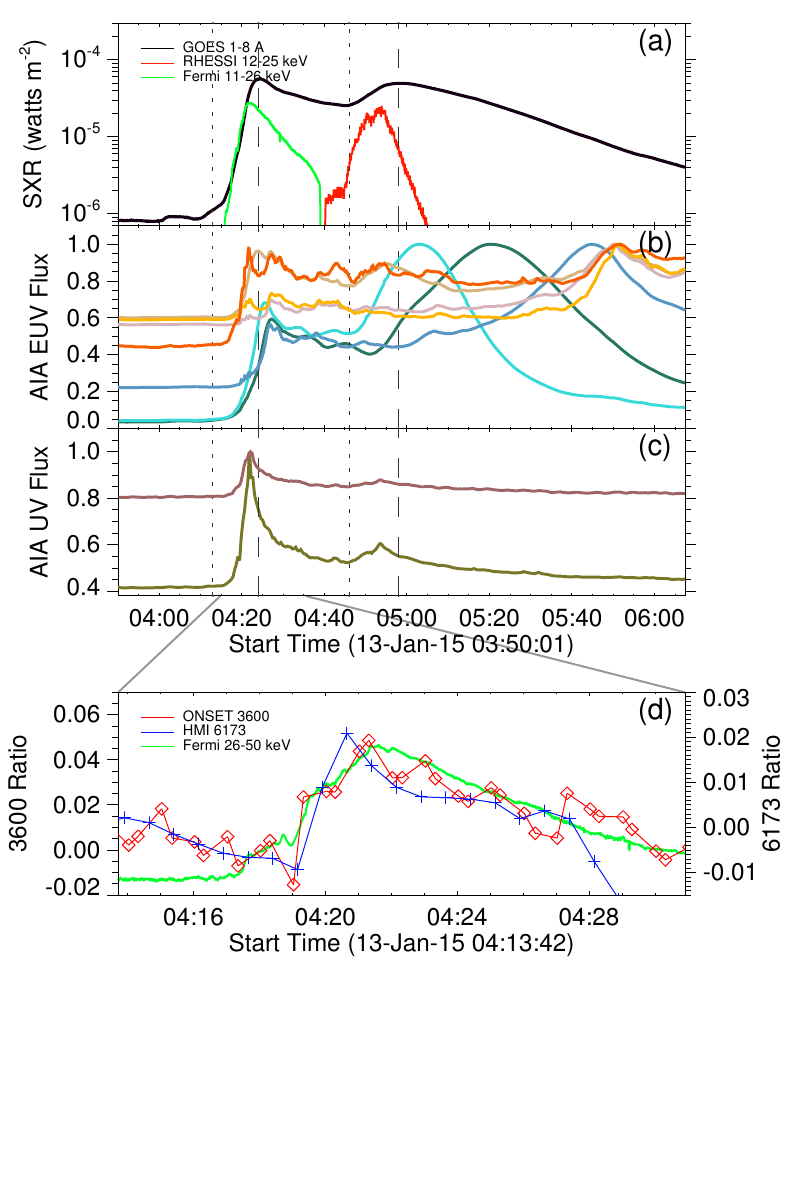} \vspace{-0.1\textwidth}}
\caption{(a) \textit{GOES} 1--8 {\AA} SXR flux (black) showing the evolution of the flare on 2015 January 13. \textit{Fermi} HXR 11-26 (green) and \textit{RHESSI} 12-25 keV flux (red) are also plotted. (b) Normalized EUV fluxes of the flare region at the AIA 131 (cyan), 94 (green), 335 (blue), 193 (orange), 211 (purple), 171 (yellow), and 304 (red) {\AA} passbands. (c) Normalized UV fluxes at the AIA 1600 (purple) and 1700 (atrovirens) {\AA} passbands. Two vertical dotted (dash) lines denote the onset (peak) times of two episodes of the flare. (d) The evolution of the WL continuum enhancement at the \textit{ONSET} 3600 (red) and HMI 6173 (blue) {\AA} passbands. The green line displays the \textit{Fermi} 26-50 keV (green) flux.}
\label{flux}
\end{figure}

It is thought that in most cases, two-ribbon flares are closely associated with the successful eruption of MFRs \citep{chen11_review}. Nevertheless, it is recently noticed that a failed MFR eruption is also able to produce a two-ribbon flare, whose characteristics are qualitatively similar to that of CME-associated flares. However, it is still unclear how much quantitative difference, in particular in the WL emission, exists in these two kinds of flares. Here, we investigate a CME-less flare that shows a continuum enhancement in the 3600 {\AA} passband. In Sections 2 and 3, we display observations of the failed eruption of the MFR and the two-ribbon flare in WL, respectively. In Section 4, we present the spectroscopic features of a flare kernel. Summary and discussion are given in Section 5.  

\section{Unsuccessful Eruption of the MFR}

\begin{figure}
\center {\includegraphics[width=5.85cm]{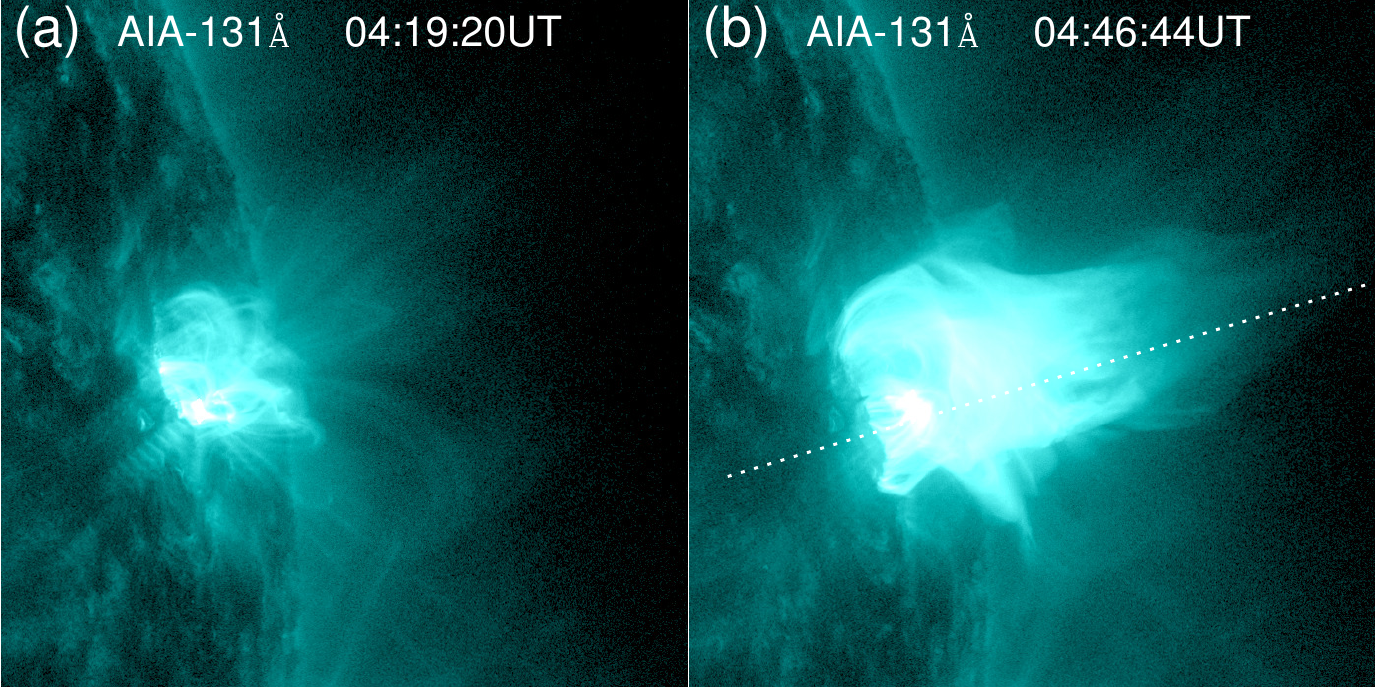}\hspace{-0.038\textwidth} \vspace{-0.03\textwidth}}
\center {\includegraphics[width=8cm]{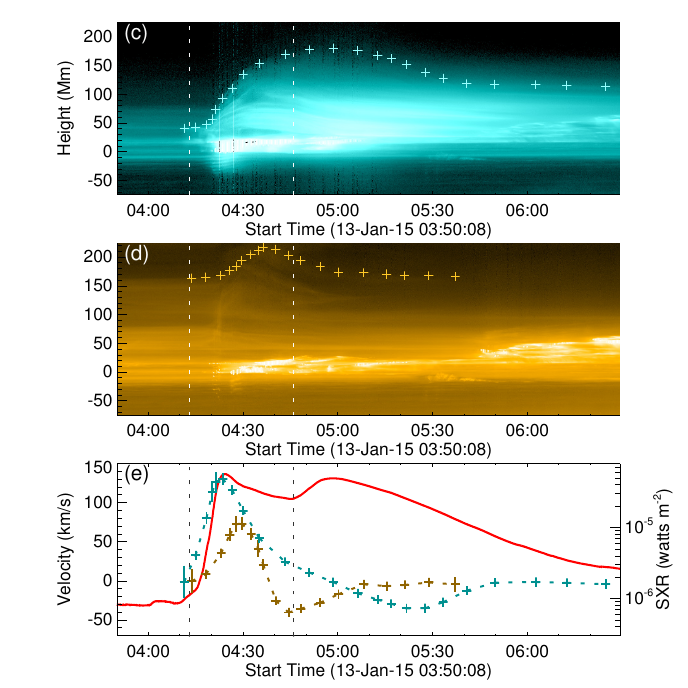}\vspace{-0.01\textwidth}}
\caption{(a) and (b) The AIA 131 {\AA} images showing the failed MFR eruption. The oblique dashed line denotes a slice that is along the direction of the MFR eruption. (c) and (d) The slice-time plots of the AIA 131 and 171 {\AA} passbands. The cyan and yellow pluses denote the height-time measurement of the MFR and overlying field, respectively. (e) The temporal variation of the velocity of the MFR (cyan) and overlying field (yellow). The \textit{GOES} 1--8 {\AA} SXR flux (red) is also plotted. The two vertical dash lines denote the onsets of two episodes of the flare, respectively.}
\label{slice}
\end{figure}

On 2015 January 13, a major flare consisting of two energy release episodes (episode I and episode II) occurred at the heliographic coordinates $\sim$N06W70. The corresponding \textit{GOES} soft X-ray (SXR) flux starts to increase at $\sim$04:13 UT and peaks at $\sim$04:24 UT. After a decay of $\sim$22 minutes, the SXR flux increases again and reaches the maximum at $\sim$04:58 UT (Figure \ref{flux}a). The two peaks in the SXR flux are also visible in the lightcurve of the total EUV and UV emissions of the flare region (Figure \ref{flux}b and \ref{flux}c). It is seen that the UV emissions at the 1600 and 1700 {\AA} passbands (Figure \ref{flux}c), which mostly originate in the lower atmosphere, mainly appears in episode I, while the EUV emissions, which are from the corona, primarily appear in episode II. 

\begin{figure*}
\center {\includegraphics[width=12cm]{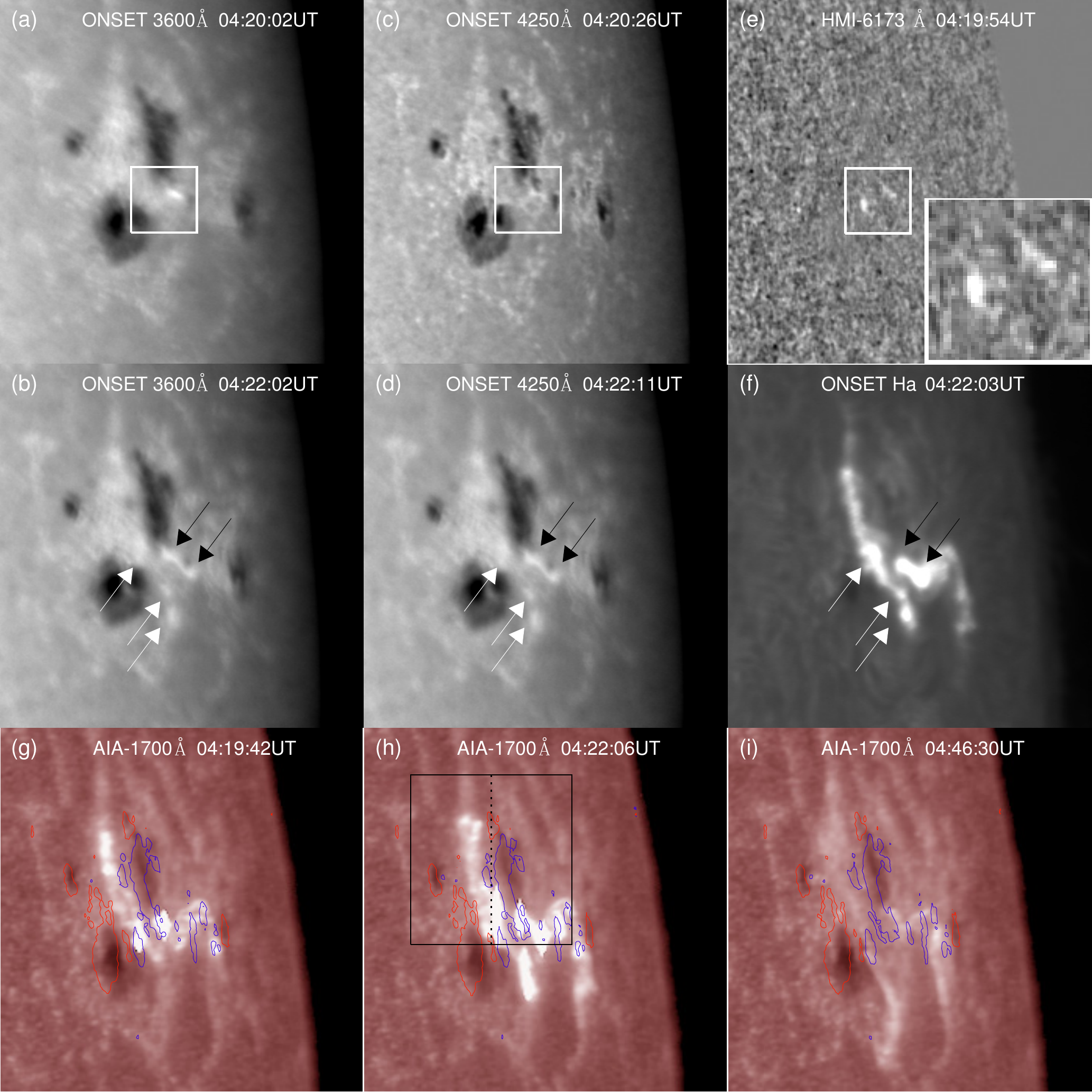}}
\caption{(a)--(f) \textit{ONSET} 3600 {\AA}, 4250 {\AA}, H$\alpha$, and HMI 6173 {\AA} images displaying the evolution of the ribbon-like WLF. The white boxes denote the primary region of the WL enhancement, which is also used for calculating the WL contrasts at 3600 {\AA} and 6173 {\AA}. The bigger white box in the panel (c) is a zooming in of the primary WL region. The white and black arrows indicate two WL flare ribbons. (g)--(i) The AIA 1700 {\AA} images showing two ribbons in the chromosphere. The red (blue) contours indicate the positive (negative) polarity of magnetic field. The black box in the panel (h) displays the field of view of \textit{IRIS}. The vertical dashed line presents the position of the slit.}
\label{onset}
\end{figure*}

The emission in episode II is mainly associated with the descending motion of a hot structure, which is most likely to be an MFR \citep{cheng11_fluxrope,patsourakos13,song14}. It commences when the erupting MFR catches up with the overlying field and starts to go down (Figure \ref{slice}a--\ref{slice}c and attached movie). The absence of a CME in the WL coronagraph LASCO/C2 also proves that the eruption of the MFR is unsuccessful. The AIA 131 and 94 {\AA} images show that the MFR starts to rise up at $\sim$04:10 UT. However, it starts to go down from $\sim$04:46 UT. At the same time, the flare emission in episode II starts to increase and the unsuccessfully erupting MFR is further heated.

We put a slice along the direction of the MFR eruption and make the slice-time plots of the AIA 131 and 171 {\AA} passbands. One can clearly see an ascending motion followed by a descending one of the MFR and overlying field (Figure \ref{slice}c and \ref{slice}d). Using the slice-time plots, we measure the heights of the MFR and overlying field (pluses in Figure \ref{slice}c and \ref{slice}d). Applying the first order numerical derivative, we calculate the velocities. The results are shown in Figure \ref{slice}e. We find that the MFR is rapidly accelerated in the rising phase of the flare episode I. The velocity increases to $\sim$150 km s$^{-1}$ at the peak of the \textit{GOES} SXR flux ($\sim$04:23 UT). After that, probably due to the strong strapping effect of the overlying field, the MFR starts to be decelerated. At $\sim$04:30 UT, the decelerating MFR and accelerating overlying field come to the same velocity ($\sim$70 km s$^{-1}$). Then, the overlying field starts to speed down. At $\sim$04:35 UT, it turns to move downward. When the descending velocity reaches the maximum ($\sim$--70 km s$^{-1}$), the flare episode II starts. Subsequently, both of the MFR and the overlying field move downward. Note that, the descending velocity of the MFR is smaller that that of the overlying field, implying that a collision between them may happen.

\section{WL Emission of the Flare Ribbons}

\begin{figure*}
%\center {\includegraphics[width=18cm, angle=90]{f_spec.pdf}}
\hspace{0.6\textwidth}
\center {\includegraphics[width=18cm]{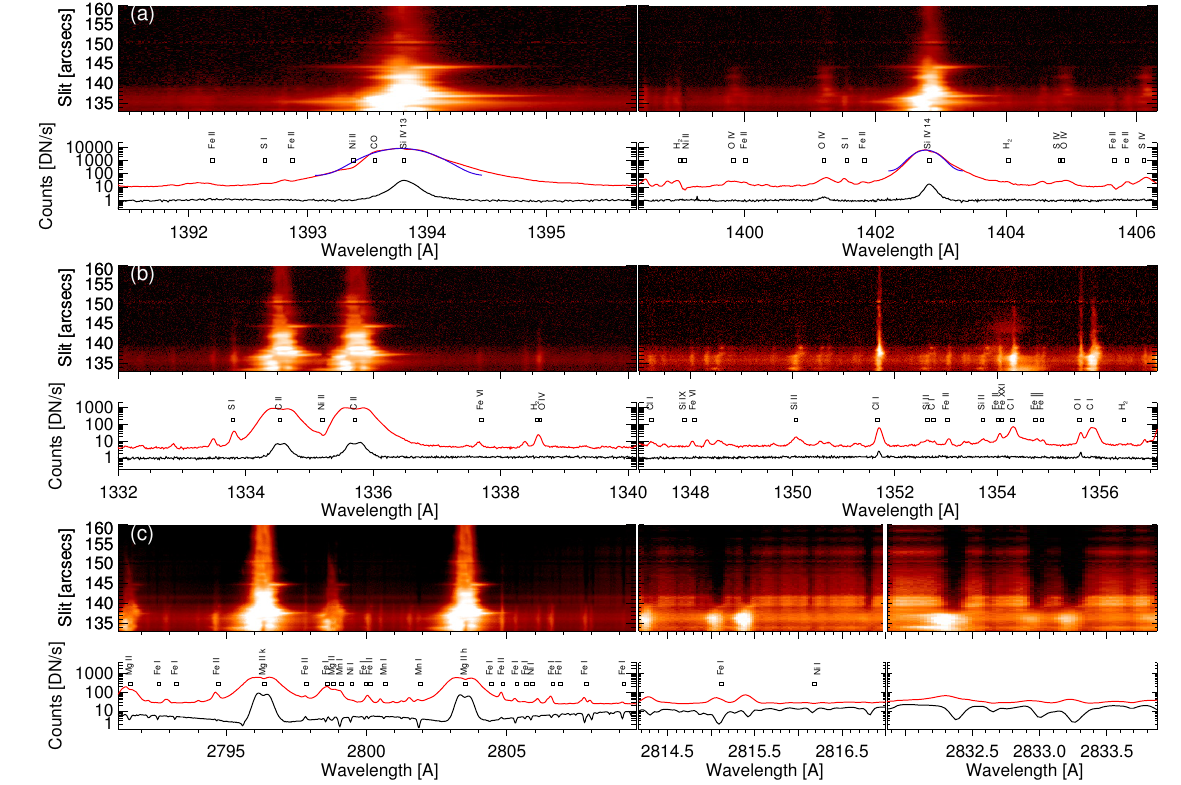}}
\caption{\textit{IRIS} FUV and NUV spectrograms (top rows) and the average profiles of the spectral lines (bottom rows) in the WL kernel (red; y=[134\arcsec,139\arcsec]) and quiet region (black; y=[160\arcsec,190\arcsec]). The two blue lines display the single Gaussian fitting to the average profiles of the \ion{Si}{4} lines.}
\label{iris}
\end{figure*}

Although the MFR eruption is not successful, there still appear two elongated ribbon-like flare brightenings even at the WL passbands that are well observed by the \textit{Optical and Near-infrared Solar Eruption Tracer} \citep[\textit{ONSET};][]{fang13} of Nanjing University with a field of view of 10 arcmin. The \textit{ONSET} is installed at a new solar observing site (E102.57$^{\circ}$, N24.38$^{\circ}$) near Fuxian Lake in Yunnan Province, southwest China, where the seeing is generally stable and better than 1\arcsec, thus guaranteeing the high quality of data \citep{hao12}. It routinely images the photosphere at 3600 {\AA} and 4250 {\AA} with a wavelength bandpass of 15 {\AA}. It also images the upper chromosphere using the H$\alpha$ (6562.8 {\AA}) line with a bandpass of 0.25 {\AA}. Generally, users can directly use the Level-1 images\footnote{http://sdac.nju.edu.cn}, which have been pre-processed with the calibration of the flat-field and dark current. Table \ref{tb1} presents the capabilities of various telescopes that observe this WLF. One can see that the \textit{ONSET} has a high temporal resolution. On 2015 January 13, the seeing near Fuxian Lake is pretty good, and the spatial resolution of \textit{ONSET} approaches the upper limit, i.e., better than 1\arcsec (the pixel size is 0.24\arcsec), which is comparable with that of the AIA.

Figure \ref{flux}d shows the evolution of the WL continuum enhancement, which is derived by averaging the intensity over the pixels showing WL enhancement with the background subtracted. We find that the evolution of the WL enhancement at the passbands 3600 {\AA} (\textit{ONSET}) and 6173 {\AA}, which is from the Helioseismic and Magnetic Imager \citep[HMI;][]{schou12} on board \textit{SDO}, is well coincident with the variation of \textit{Fermi} HXR 26--50 keV flux.

Moreover, we find an evolution of the WL brightening from the two isolated kernels to two well-shaped ribbons. At $\sim$04:20 UT, 7 minutes after the flare occurrence, two WL kernels appear at the edge of the sunspot penumbra (Figure \ref{onset}a and \ref{onset}b). The base-difference image of the HMI 6173.34 {\AA} continuum intensity also confirms the appearance of the two WL kernels (Figure \ref{onset}e). With the flare development, the left WL kernel extends toward the north, while the right one extends toward the southwest. At $\sim$04:22 UT, the WL brightening makes up two well-shaped ribbon-like structures, which are almost parallel to the polarity inversion line (PIL) of the active region (Figure \ref{onset}c and \ref{onset}d). As compared with two elongated ribbons in H$\alpha$ and 1700 {\AA} (Figure \ref{onset}f--\ref{onset}i), the WL ribbons are shorter and thinner. These results indicate that this flare is a type I WLF, and the increase of the WL emission is associated with high-energy electrons, which are accelerated by the flare reconnection in the corona and then may sequentially heat the chromoshpere and even below along the direction of the PIL, finally forming two well-shaped WL ribbons.

\begin{figure*}
\center {\includegraphics[width=17cm]{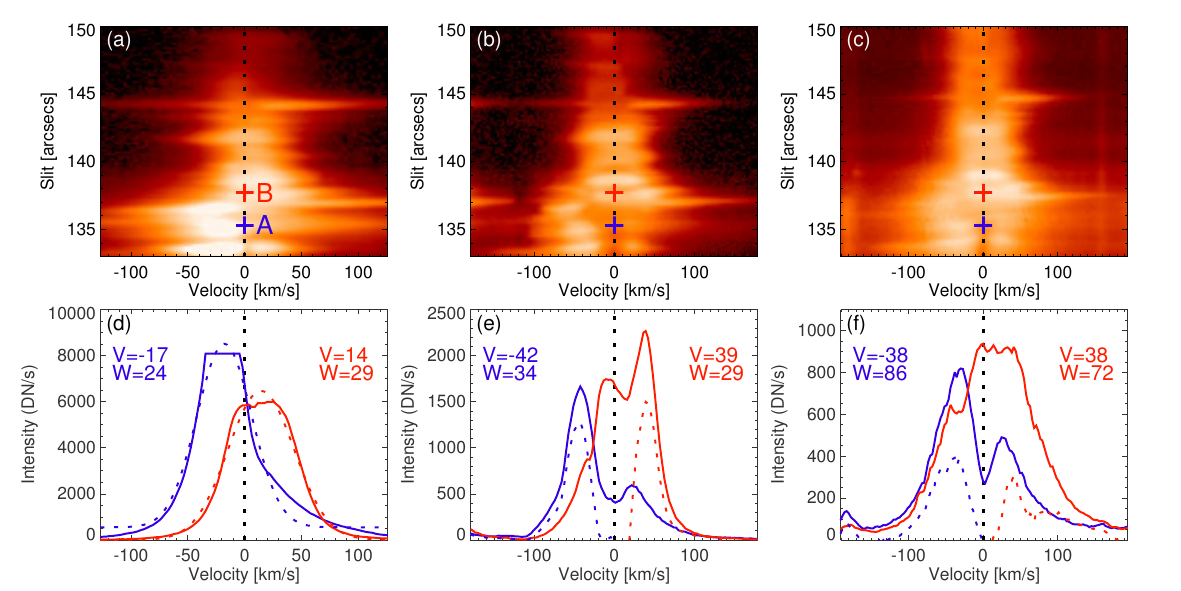}}
\caption{(a)--(c) Spectrograms of the \ion{Si}{4} 1402.77 {\AA}, \ion{C}{2} 1335.7077 {\AA}, and \ion{Mg}{2} k 2796.347 {\AA} lines in the WL kernel (y=[133\arcsec,150\arcsec]). The vertical dotted lines show the line centers. The blue and red pluses denote the two adjacent pixels in the WL kernel. (d)--(f) Profiles of the \ion{Si}{4}, \ion{C}{2}, and \ion{Mg}{2} k lines at the blue and red pixels. The dotted curves in panel (d) display the single Gaussian fittings to the profiles of the \ion{Si}{4} line, that in panels (e) and (f) show the differences between the blue wings and the red wings of the \ion{C}{2} and \ion{Mg}{2} k lines.}
\label{profile}
\end{figure*}

\section{Spectroscopic Analysis of the Flare Ribbons}
The continuum enhancement of the flare is also observed by the recently launched \textit{Interface Region Imaging Spectrograph} \citep[\textit{IRIS};][]{depontieu14}, which acquires spectra in the passbands 1332--1358 {\AA} (FUV1), 1389--1407 {\AA} (FUV2), and 2783--2834 {\AA} (NUV) that include three doublet bright lines, the \ion{C}{2} 1334/1335 ($\sim$$10^{4.3}$ K) and \ion{Si}{4} 1394/1403 ($\sim$$10^{4.8}$ K) lines forming in the transition region and the \ion{Mg}{2} h/k ($\sim$$10^{4}$ K) line in the chromosphere. At $\sim$04:22 UT, the slit of \textit{IRIS} is positioned crossing the left WL kernel (Figure \ref{onset}h). The corresponding spectrogram and average profiles of the spectral lines at the WL kernel are displayed in Figure \ref{iris}, together with the profiles at the quiet region for comparison. It can be clearly seen that the continuum enhancement appears at the whole FUV (NUV) passbands as expected. At the FUV (NUV) passbands, the continuum enhancement is as large as $\sim$10 (5) times the background, respectively.

Besides the continuum enhancement, \textit{IRIS} also discloses some spectral features of the WLF (Figure \ref{iris}a--\ref{iris}c). First, the average profiles of the \ion{Si}{4}, \ion{C}{2}, and \ion{Mg}{2} h/k lines display an extremely large peak enhancement and significant non-thermal broadening (bottom panel of Figure \ref{iris}a). By applying the single Gaussian fitting, the non-thermal velocity of the \ion{Si}{4} line is found to be 40--60 km s$^{-1}$. Second, those lines like H$_{2}$ 1338.57 {\AA}, Cl I 1347.2397 {\AA}, and Fe II 1353.023 {\AA} etc \citep{kelly87}, that are very weak and even invisible at the quiet region, get obviously enhanced when the WL enhancement appears. Third, some absorption lines near the far wing of \ion{Mg}{2} h/k, which are thought to be possibly from the lower chromosphere and photosphere with the formation temperature of $\sim$$10^{3.7-3.8}$ K \citep{depontieu14}, turn to emission. These observations also support the scenario of high-energy electron bombardment, which heat the upper chromosphere and lead to the FUV/NUV enhancement and non-thermal line broadening. Change of the lines from absorption to emission implies a heating even lower in the lower chromosphere and photosphere. High-energy electrons cannot precipitate directly to these layers, which are most probably heated through the radiative backwarming effect \citep{machado89,metcalf90,ding03,liuying01}

Carefully inspecting the spectrogram in the WL kernel, we further find that the Doppler shift pattern is quite location-dependent: at the strongly bright (saturated) pixels, the \ion{Si}{4}, \ion{C}{2}, and \ion{Mg}{2} h/k lines mostly display a blueshift; in contrast, at the moderately bright pixels, the lines are redshifted (Figure \ref{profile}a--\ref{profile}c). Usually, the blueshifts are thought to be produced by chromospheric evaporation, while the redshifts are caused by chromospheric condensation (downward moving compression front). The time scale and magnitudes of both upward and downward velocities depend on the energy input rate and also the atmospheric environment. The facts that adjacent pixels witnessing different dynamic phenomena can mainly be due to the different energy injection rates, as reflected by the different line intensities. Note that, the centers of the \ion{C}{2} and \ion{Si}{4} lines are calibrated by assuming that the nearby \ion{O}{1} 1355.5977 and \ion{Fe}{2} 1405.608 {\AA} lines have a zero velocity. The reference center of the \ion{Mg}{2} line is taken as that of the average profile in the quiet region \citep{cheng15_formation}. 

Furthermore, we quantify the Doppler and non-thermal velocity at two adjacent pixels (A and B) in the WL kernel. Pixel A corresponds to the strongly bright region, and B is at the moderately bright region. The profiles of the \ion{Si}{4} 1402.77 {\AA}, \ion{C}{2} 1335.7077 {\AA}, \ion{Mg}{2} k 2796.347 {\AA} lines at A and B are shown in Figure \ref{profile}d--\ref{profile}f. One can see that the profiles of the \ion{Si}{4} line are nearly Gaussian. The most obvious difference is that the center of the profile at A is blueshifted, while that at B is redshifted. The single Gaussian fitting gives the Doppler velocity of --17 and 14 km s$^{-1}$ and non-thermal velocity of 23 and 28 km s$^{-1}$ for A and B, respectively. As for the \ion{C}{2} and \ion{Mg}{2} k lines, the profiles generally deviate from Gaussian and show a self-absorption in the line center due to the optically thick effect of the lines \citep{iris_leenaarts2,iris_leenaarts1}. It is usually hard to get a velocity value causing the line asymmetries. A simple and approximate way is to subtract the relatively less perturbed wing from the more perturbed wing. The shift of the peak in the residual profile is then regarded as the velocity of the plasma producing the excess emission \citep{cheng15_formation}. In this way, the Doppler velocities of the \ion{C}{2} and \ion{Mg}{2} k lines are --42 and --38 km s$^{-1}$ at A, while 39 and 38 km s$^{-1}$ at B, respectively. 

\section{Summary and Discussion}
In this Letter, we investigate a CME-less two ribbon-like WLF that experiences two episodes of energy release. The first episode begins with the eruption of the MFR and produces a ribbon-like WL brightening; while the second one starts when the MFR catches up with the overlying field. The two energy release episodes provide a possible scenario that is related to the origin of the later phase of flares, which is defined as a second emission peak in the warm coronal lines \citep[e.g., Fe XVI 33.5 nm;][]{woods11}. In the second energy release episode, the reconnection can continuously release energy for plasma heating, though in a lower rate, and produce the EUV late phase \citep[e.g.,][]{dai13,sunxd13,liying14,kai15}. Energy release in episode II is also evidenced by appearance of a big fire ball with a temperature up to $\sim$10 MK corresponding to the unsuccessfully erupted MFR \citep[also see;][]{song14}. 

A two-ribbon flare is usually believed only associated with a successful eruption of the MFR or CME. Nevertheless, from this event, we find that a failed eruption is also able to generate a two-ribbon flare even at the WL passband. Generally speaking, events of such a kind are very few in observations. However, it is noticed that a recent active region (NOAA 12192) produces tens of two-ribbon flares, but none of them is associated with the CME. Two possible reasons have been proposed for CME-less flares: one is that the non-potentiality of the erupting core field is too weak, the other is that the corresponding overlying field is too strong \citep{ji03,guo10_index,cheng11_flare,sun15,thalmann15,chenhd15}.

Moreover, the two-ribbon flare is also visible at 3600 {\AA}. Comparing with the HMI, which only observes the two isolated kernels, the \textit{ONSET} successfully captures not only the two isolated kernels but also their development into two elongated ribbons in detail thanks to the high temporal and good enough spatial resolution. The evolution of the WL continuum emission is almost synchronous with the variation of \textit{Fermi} HXR 26--50 keV flux. The \textit{IRIS} observations of a WL kernel reveal an apparent increase of the continuum emission at the whole FUV (NUV) passbands, as well as an extremely large enhancement and significant non-thermal broadening of the \ion{Si}{4}, \ion{C}{2}, and \ion{Mg}{2} h/k lines. Many weak lines, including some absorption lines before the flare, also present a strong emission when the flare occurs. These results can be interpreted in terms of the well-adopted scenario: the flare reconnection produces high-energy electrons that heat the chromosphere, giving rise to the enhancement of the FUV (NUV) continuum emission, as well as non-thermal line broadening of the lines forming there. However, because the electrons with energies $<$ 200 keV are hardly possible to reach and directly heat the lower chromosphere and even the photosphere \citep{vernazza81,ding03}, an indirect heating through radiative backwarming is usually used for interpreting the WL continuum enhancement and the photospheric absorption lines turning to emission.

Owing to the unprecedented high spatial and spectral resolution of \textit{IRIS}, we find that the Doppler shift pattern of the spectral lines is quite location-dependent. At the strongly bright pixels, all lines are prone to be blueshifted; in contrast, at the moderately bright pixels, they tend to be redshifted. This could be due to the different heating rates in different places that produce different magnitudes and time scales, and possibly even different signs, of the velocities in the chromospheric evaporation (condensation) process. Different patterns of evaporations have also been reported by \citet{liying11} for, however, well separated flare kernels. Previous observations also showed that for a specific site, the hotter coronal lines are usually blueshifted while the cooler chromospheric and transition region lines are in more cases redshifted \citep{tian14_reconnection,tian15}, and that there appears a transition from blueshift to redshift for the same line with the flare development \citep{milligan09,chen10_eis}

\acknowledgements We are grateful to the referee for his/her comments that helped improve the manuscript. We also thank Hui Tian for useful discussions and Chuan Li for providing \textit{Fermi} data. \textit{SDO} is a mission of NASAs Living With a Star Program. IRIS is a NASA small explorer mission developed and operated by LMSAL with mission operations executed at NASA Ames Research center and major contributions to downlink communications funded by the Norwegian Space Center (NSC, Norway) through an ESA PRODEX contract. This work is supported by NSFC under grants 11303016, 11373023, and NKBRSF under grants 2011CB811402 and 2014CB744203, as well as the Specialized Research Fund for State Key Laboratories.

\end{document}